\documentclass[aps,prd,preprint,tightenlines,superscriptaddress,showpacs,byrevtex]{revtex4}
\usepackage{graphicx}

\draft 

\def\BB   {B\bar{B}}
\def\qqbar{q\bar{q}}

\def\BtoXsgamma{B\to X_s\gamma}
\def\Xs{X_s}
\def\BtoKll{B\to K\ell^+\ell^-}
\def\BtoKstarll{B\to K^*\ell^+\ell^-}

\def\BtoXsll{B\to\Xs\ell^+\ell^-}
\def\BtoXsee{B\to\Xs e^+e^-}
\def\BtoXsmumu{B\to\Xs\mu^+\mu^-}
\def\BtoXsemu{B\to\Xs e^\pm\mu^\mp}
\def\BtoJpsiX{B\to J/\psi X}

\def\GeV{{\rm~GeV}}
\def\GeVc{{\rm~GeV}/c}
\def\GeVcc{{\rm~GeV}/c^2}
\def\MeV{{\rm~MeV}}

\def\MeVcc{{\rm~MeV}/c^2}
\def\cm{{\rm~cm}}
\def\fbinv{{\rm~fb}^{-1}}

\def\Mbc{M_{\rm bc}}

\def\DeltaE{\Delta E}

\def\Mll{M_{\ell^+\ell^-}}

\def\Mxs{M_{X_s}}

\def\Br{{\cal B}}
\def\calL{{\cal L}}

\def\Fsfw{{\cal F}_{\rm FW}}
\def\Fmiss{{\cal F}_{\rm miss}}

\def\EffElectron{94.3\%}
\def\EffMuon{93.2\%}
\def\FakeElectron{0.1\%}
\def\FakeMuon{0.9\%}
\def\EffKaon{90\%}
\def\FakeKaon{6\%}

\def\EffBtoXsee{   2.59\,\pm0.20\,^{+0.45}_{-0.42}}
\def\EffBtoXsmumu{ 2.89\,\pm0.24\,^{+0.52}_{-0.49}}
\def\EffBtoXsll{   2.74\,\pm0.22\,^{+0.48}_{-0.45}}

\def\BrBtoXsee{   4.04\,\pm1.30\,^{+0.87}_{-0.83}}    
\def\BrBtoXsmumu{ 4.13\,\pm1.05\,^{+0.85}_{-0.81}}  
\def\BrBtoXsll{   4.11\,\pm0.83\,^{+0.85}_{-0.81}}    
\def\BrBtoXsllFull{4.11\,\pm 0.83({\rm stat})\,^{+0.85}_{-0.81}({\rm syst})}

\def\SigBtoXsee{   3.2}
\def\SigBtoXsmumu{ 4.4}
\def\SigBtoXsll{   5.4}

\begin{document}

\renewcommand{\thefootnote}{\fnsymbol{footnote}}

\vspace*{-3\baselineskip}
\resizebox{!}{2.8cm}{\includegraphics{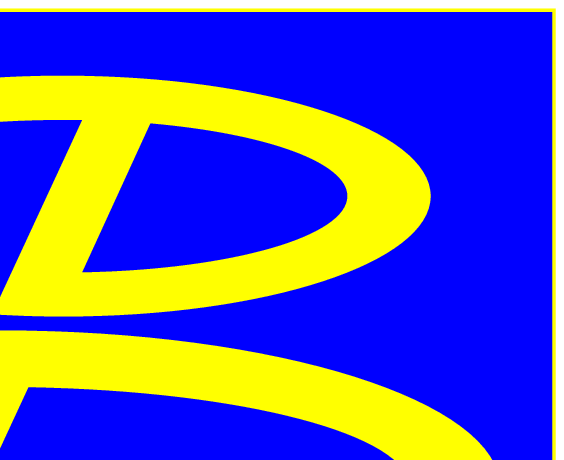}}
\preprint{\vbox{\vspace*{2cm}
		  \hbox{KEK   Preprint 2005-5}
		  \hbox{Belle Prerpint 2005-14}
}}
\vspace*{0.5cm}

\title{
Improved Measurement of the Electroweak Penguin Process            
\boldmath$\BtoXsll$  
}

\affiliation{Budker Institute of Nuclear Physics, Novosibirsk}
\affiliation{Chiba University, Chiba}
\affiliation{Chonnam National University, Kwangju}
\affiliation{University of Cincinnati, Cincinnati, Ohio 45221}
\affiliation{University of Frankfurt, Frankfurt}
\affiliation{Gyeongsang National University, Chinju}
\affiliation{University of Hawaii, Honolulu, Hawaii 96822}
\affiliation{High Energy Accelerator Research Organization (KEK), Tsukuba}
\affiliation{Hiroshima Institute of Technology, Hiroshima}
\affiliation{Institute of High Energy Physics, Chinese Academy of Sciences, Beijing}
\affiliation{Institute of High Energy Physics, Vienna}
\affiliation{Institute for Theoretical and Experimental Physics, Moscow}
\affiliation{J. Stefan Institute, Ljubljana}
\affiliation{Kanagawa University, Yokohama}
\affiliation{Korea University, Seoul}
\affiliation{Kyungpook National University, Taegu}
\affiliation{Swiss Federal Institute of Technology of Lausanne, EPFL, Lausanne}
\affiliation{University of Ljubljana, Ljubljana}
\affiliation{University of Maribor, Maribor}
\affiliation{University of Melbourne, Victoria}
\affiliation{Nagoya University, Nagoya}
\affiliation{Nara Women's University, Nara}
\affiliation{National Central University, Chung-li}
\affiliation{National United University, Miao Li}
\affiliation{Department of Physics, National Taiwan University, Taipei}
\affiliation{H. Niewodniczanski Institute of Nuclear Physics, Krakow}
\affiliation{Nihon Dental College, Niigata}
\affiliation{Niigata University, Niigata}
\affiliation{Osaka City University, Osaka}
\affiliation{Osaka University, Osaka}
\affiliation{Panjab University, Chandigarh}
\affiliation{Peking University, Beijing}
\affiliation{Princeton University, Princeton, New Jersey 08544}
\affiliation{Saga University, Saga}
\affiliation{University of Science and Technology of China, Hefei}
\affiliation{Seoul National University, Seoul}
\affiliation{Sungkyunkwan University, Suwon}
\affiliation{University of Sydney, Sydney NSW}
\affiliation{Tata Institute of Fundamental Research, Bombay}
\affiliation{Toho University, Funabashi}
\affiliation{Tohoku Gakuin University, Tagajo}
\affiliation{Tohoku University, Sendai}
\affiliation{Department of Physics, University of Tokyo, Tokyo}
\affiliation{Tokyo Institute of Technology, Tokyo}
\affiliation{Tokyo Metropolitan University, Tokyo}
\affiliation{Tokyo University of Agriculture and Technology, Tokyo}
\affiliation{University of Tsukuba, Tsukuba}
\affiliation{Virginia Polytechnic Institute and State University, Blacksburg, Virginia 24061}
\affiliation{Yonsei University, Seoul}
 \author{M.~Iwasaki}\affiliation{Department of Physics, University of Tokyo, Tokyo} 
 \author{K.~Itoh}\affiliation{Department of Physics, University of Tokyo, Tokyo} 
 \author{H.~Aihara}\affiliation{Department of Physics, University of Tokyo, Tokyo} 
   \author{K.~Abe}\affiliation{High Energy Accelerator Research Organization (KEK), Tsukuba} 
   \author{K.~Abe}\affiliation{Tohoku Gakuin University, Tagajo} 
   \author{I.~Adachi}\affiliation{High Energy Accelerator Research Organization (KEK), Tsukuba} 
   \author{Y.~Asano}\affiliation{University of Tsukuba, Tsukuba} 
   \author{T.~Aushev}\affiliation{Institute for Theoretical and Experimental Physics, Moscow} 
   \author{S.~Bahinipati}\affiliation{University of Cincinnati, Cincinnati, Ohio 45221} 
   \author{A.~M.~Bakich}\affiliation{University of Sydney, Sydney NSW} 
% ...
\collaboration{The Belle Collaboration}

\begin{abstract}
We present an improved measurement of the branching fraction for the
electroweak penguin process $\BtoXsll$,
where $\ell$ is an electron or a muon
and $\Xs$ is a hadronic system containing an $s$-quark.
The measurement is based on a sample of 
$152 \times 10^{6}$~$\Upsilon(4S) \to \BB$ events
collected with the Belle detector at the KEKB 
energy asymmetric $e^+e^-$ collider.
The $\Xs$ hadronic system is reconstructed 
from one $K^{\pm}$ or $K^{0}_{\rm S}$ and up to 
four pions, where at most one pion can be neutral.
Averaging over both lepton flavors, 
the inclusive branching fraction is measured to be
$\Br(\BtoXsll)=(\BrBtoXsllFull)\times10^{-6}$
for $M_{\ell^{+} \ell^{-}} > 0.2\GeVcc$.

\end{abstract}




\maketitle
 


\section{Introduction}
\label{sec:Introduction}

In the Standard Model (SM), 
the rare decay $\BtoXsll$ proceeds through a 
$b \to s \ell^{+} \ell^{-}$ transition, 
which is forbidden at tree level.
Such a flavor-changing neutral current (FCNC)
process can occur at higher order via electroweak penguin and $W^{+} W^{-}$
box diagrams.
The $b \to s \ell^{+} \ell^{-}$ transition therefore allows 
deeper insight into the effective Hamiltonian that describes FCNC processes
and is sensitive to the effects of non-SM physics that may enter these
loops; see, for example, Refs~\cite{Ali02,Hurth03}.

Recent SM calculations
of the inclusive $\BtoXsll$ branching fractions 
for $M_{l^{+} l^{-}} > 0.2$ GeV/$c^2$
predict
${\cal B}(\BtoXsee) = 
(4.2 \pm 0.7) \times 10^{-6}$~\cite{Ali02,Ali_ICHEP02}, 
and 
${\cal B}(\BtoXsll) = (4.6 \pm 0.8) \times 10^{-6}$~\cite{Isidori}.
Both the Belle and BaBar collaborations have 
observed exclusive 
$B \to K \ell^{+} \ell^{-}$  and 
$K^{*} \ell^{+} \ell^{-}$
decays~\cite{Belle_Kll,Belle_Kstll,BaBar_Kstll}
and have measured the rate for inclusive $\BtoXsll$ decay~\cite{Belle_sll,BaBar_sll}.

In this analysis, we study the inclusive $\BtoXsll$
process by semi-inclusively reconstructing 
the final state from a pair of electrons
or muons and a hadronic system consisting of one $K^{\pm}$ or
$K^{0}_{\rm S}$ 
and up to four pions, where at most one pion can be neutral.
This semi-inclusive reconstruction approach~\cite{CLEO_bsgamma95}
allows approximately 53\% of the full inclusive rate to be reconstructed.
If the contribution of the modes containing a $K^0_{\rm L}$ 
is taken to be equal to that containing a $K^0_{\rm S}$,
the missing states that remain unaccounted for represent $\sim$30\%
of the total rate.
We require the hadronic mass for the selected final states
to be less than 2.0$\GeVcc$ to reduce combinatorial background.
We correct for the missing modes and the effect of the 
hadronic mass requirement to extract the 
inclusive $\BtoXsll$ decay rate for $M_{\ell^{+} \ell^{-}} > 0.2\GeVcc$.
This measurement updates and supersedes 
our previous result~\cite{Belle_sll}, 
which was based on a sample of $65 \times 10^{6}$~$\BB$ pairs.

\section{The Belle Detector and Data Sample}

We use a data sample collected on the $\Upsilon(4S)$ resonance with the Belle
detector at the KEKB 
energy asymmetric $e^+e^-$ collider
(3.5 GeV on 8 GeV) \cite{bib:kekb-nim}.  This sample comprises
$152 \times 10^{6}$ $B$ meson pairs, 
corresponding to an integrated luminosity of $140~\fbinv$.
%
%
A detailed description of the Belle detector can be found 
elsewhere~\cite{bib:belle-nim}.
A three-layer silicon vertex detector (SVD) and 
a 50-layer central drift chamber (CDC) are used for
tracking and identification of charged particles. 
An array of aerogel threshold \v{C}erenkov
counters (ACC) and  time-of-flight scintillation counters (TOF)
are also used for charged particle identification.
An electromagnetic calorimeter comprised of Tl-doped CsI crystals
(ECL) measures the energy of electromagnetic particles
and is also used for electron identification.
These detectors are located inside a superconducting 
solenoid coil that provides a 1.5~T magnetic field.
An iron flux-return located outside of the coil is instrumented
with resistive plate counters to identify muons (KLM).

Particle identification for 
$e^{\pm}$, $\mu^{\pm}$, $K^{\pm}$, $K^{0}_{\rm S}$, $\pi^{\pm}$ and $\pi^{0}$
is important for this analysis. 
Electron identification is based on 
the ratio of the cluster energy to the track momentum ($E/p$), 
the specific energy loss ($dE/dx$) in the CDC, 
the position and shower shape of the cluster in the ECL and 
the response from the ACC.
Muon identification is based on the
hit positions and the depth of penetration into the ECL and KLM.
Electrons and muons are required to have laboratory-frame momenta
greater than $0.4\GeVc$ and $0.8\GeVc$, respectively. 
To select good muon candidates, we apply a kaon veto.
We find an electron (muon)
selection efficiency of $\EffElectron$ ($\EffMuon$) with a
$\FakeElectron$ ($\FakeMuon$) pion to electron (muon)
mis-identification probability.
Bremsstrahlung photons from electrons are recovered by
combining each electron with photons
within a small angular region around the electron direction.
Charged kaon candidates are selected by
using information from the ACC, TOF and CDC.  
The kaon
selection efficiency is $\EffKaon$ with a pion to kaon
mis-identification probability of $\FakeKaon$.
After selecting the electron, muon and charged kaon candidate
tracks, the remaining tracks are assumed to be 
charged pions.
$K^{0}_{\rm S}$ candidates are reconstructed from pairs 
of oppositely-charged tracks with 
$|M_{\pi^{+}\pi^{-}} - M_{K^{0}_{\rm S}}| < 15\MeVcc$ 
$(\sim\!6\sigma_{M_{K^0_{\rm S}}})$.
We impose additional $K^0_{\rm S}$ selection criteria based on the
distance and the direction of the $K^0_{\rm S}$ vertex, 
and on the impact parameters of the daughter tracks.
We require all charged tracks, except those used
in the $K^0_{\rm S}$ reconstruction, to have impact parameters with respect to
the nominal interaction point of less than $1.0\cm$ in the radial
direction and $5.0\cm$ along the beam direction.
Neutral pions are required to have a laboratory-frame energy greater than $400\MeV$, 
photon daughter energies greater than $50\MeV$,
and a $\gamma \gamma$ invariant mass satisfying 
$|M_{\gamma \gamma} - M_{\pi^{0}}| < 10\MeVcc$
$(\sim\!3\sigma_{M_{\pi^0}})$.

\section{Analysis overview}
\label{sec:overview}
We reconstruct inclusive $\BtoXsll$ decays 
with a semi-inclusive reconstruction technique from a pair of electrons
or muons and one of 18 different hadronic states.
Here the hadronic system consists of one $K^{\pm}$ or $K^{0}_{\rm S}$ 
and up to four pions (at most one pion can be neutral).
Compared to a fully inclusive approach, this method has the advantage
of having strong kinematical discrimination against background
by using the 
beam-energy constrained mass 
$\Mbc=\sqrt{E_{\rm beam}^2 - p_{B}^2}$
and the energy difference $\DeltaE = E_{B} - E_{\rm beam}$,
where $E_{\rm beam}$ is the beam energy
and $E_{B}$ ($p_{B}$) is the 
reconstructed $B$ meson energy (3-momentum).
All quantities are evaluated in the $e^{+}e^{-}$ center-of-mass system (CM). 

Further background suppression
to reduce the large combinatorial backgrounds is necessary. 
The main contribution to the combinatorial background comes
from pairs of semileptonic $b \to c$ decays in $\BB$ events.
In these events, $\BtoXsll$ candidates are reconstructed with
the decay products from both $\BB$ mesons.
This background has a significant amount of missing energy
due to the neutrinos from the semileptonic decays.
Another contribution to the combinatorial background comes from
continuum events, which are efficiently suppressed 
with event-shape variables.

There are two background sources that can peak in $\Mbc$ and
$\DeltaE$.
The first comes  from
the charmonium decays $B \to J/\psi X$ and $B \to \psi(2S) X$ with
$J/\psi (\psi(2S)) \to \ell^{+} \ell^{-}$.
This charmonium background is efficiently removed with requirements on the 
dilepton mass $M_{\ell^{+} \ell^{-}}$. The resulting veto sample provides
a large control sample of decays with a signature identical to that
of the signal.
The second comes from $B \to K^{\pm} (K^0_{\rm S}) n \pi$~$(n > 1)$ 
decays with misidentification of two charged pions as leptons.
We estimate these peaking background contaminations, then
subtract them from the signal yield.

For the $\BtoXsll$ event simulation,
we use EVTGEN~\cite{EVTGEN} as the event generator,
JETSET~\cite{JETSET} to hadronize the system consisting of a
 strange quark and a spectator quark, 
and GEANT~\cite{GEANT3} for the detector simulation.
In the event generation, $\BtoXsll$ events 
are produced with a combination of 
exclusive and inclusive models.
In the hadronic mass region $M_{X_{s}} < 1.1\GeVcc$, 
exclusive $B \to K^{(*)} \ell^{+} \ell^{-}$ decays are generated 
according to Refs~\cite{Ali02,Ali00},
where the relevant form factors are computed 
using light-cone QCD sum rules. 
In the region $M_{X_{s}} > 1.1\GeVcc$, event generation is based
on a non-resonant model following Refs~\cite{Ali02,Kruger96} 
and the Fermi motion model of Ref.~\cite{Ali79}.

\section{Event selection}
\label{sec:selection}

Events are required to have a well determined primary vertex,
be tagged as multi-hadron events,
and contain 
two electrons (or muons) 
having laboratory-frame momenta greater than 
$0.4\GeVc$ (or $0.8\GeVc$ for muons).
Dilepton candidates consist of $e^{+} e^{-}$ or
$\mu^{+} \mu^{-}$ pairs. 
The two leptons are required to originate from a common vertex
and satisfy the requirement $|\Delta z| < 0.015$~cm. 
Here, $\Delta z$ is the distance between the two leptons 
along the beam direction; the $z$-coordinate 
of each lepton is determined at the point of closest approach to the beam axis.

Charmonium backgrounds are reduced by
removing $B$~candidates with a dilepton mass in the ranges
$-0.40\GeVcc < M_{ee(\gamma)}-M_{J/\psi}   < 0.15\GeVcc$,
$-0.25\GeVcc < M_{\mu\mu}    -M_{J/\psi}   < 0.10\GeVcc$,
$-0.25\GeVcc < M_{ee(\gamma)}-M_{\psi(2S)} < 0.10\GeVcc$, and
$-0.15\GeVcc < M_{\mu\mu}    -M_{\psi(2S)} < 0.10\GeVcc$.
If one of the electrons from a $J/\psi$ or $\psi(2S)$ decay erroneously 
picks up a random photon in the bremsstrahlung-recovery process,
the dilepton mass can increase sufficiently to evade
the above requirements.
Therefore the charmonium veto is applied to the dilepton mass 
before and after bremsstrahlung recovery.
Using the simulation, we estimate the remaining peaking 
charmonium background to be $1.20 \pm 0.28$ events and
$1.33 \pm 0.21$ events for the $e^{+} e^{-}$ and $\mu^{+} \mu^{-}$
channels, respectively.

The requirement $M_{l^{+}l^{-}} > 0.2\GeVcc$ is applied. 
For the $e^+e^-$ channel, this removes potential peaking background
from the conversion of photons from radiative 
$B \to X_{s} \gamma$ decays and from $\pi^0$ 
Dalitz decays.

Starting from a $\ell^{+} \ell^{-}$ pair,
$B \to X_{s} \ell^{+} \ell^{-}$ candidates are formed 
by adding either a $K^{\pm}$ or a $K^{0}_{\rm S}$ and up to
four pions, but no more than one $\pi^{0}$. 
In this manner, eighteen different hadronic topologies
are reconstructed:
$K^{\pm}$, 
$K^{\pm} \pi^0$, 
$K^{\pm} \pi^{\mp}$, 
$K^{\pm} \pi^{\mp} \pi^0$, 
$K^{\pm} \pi^{\mp} \pi^{\pm}$,
$K^{\pm} \pi^{\mp} \pi^{\pm} \pi^0$, 
$K^{\pm} \pi^{\mp} \pi^{\pm} \pi^{\mp}$,
$K^{\pm} \pi^{\mp} \pi^{\pm} \pi^{\mp} \pi^0$, 
$K^{\pm} \pi^{\mp} \pi^{\pm} \pi^{\mp} \pi^{\pm}$,
$K^0_{\rm S}$, 
$K^0_{\rm S} \pi^0$, 
$K^0_{\rm S} \pi^{\pm}$, 
$K^0_{\rm S} \pi^{\pm} \pi^0$, 
$K^0_{\rm S} \pi^{\pm} \pi^{\mp}$,
$K^0_{\rm S} \pi^{\pm} \pi^{\mp} \pi^0$,
$K^0_{\rm S} \pi^{\pm} \pi^{\mp} \pi^{\mp}$,
$K^0_{\rm S} \pi^{\pm} \pi^{\mp} \pi^{\mp} \pi^0$, and 
$K^0_{\rm S} \pi^{\pm} \pi^{\mp} \pi^{\mp} \pi^{\pm}$.

After forming the $B \to X_{s} \ell^{+} \ell^{-}$ candidates,
we apply additional requirements to suppress background.
The largest background sources are random combinations 
from continuum $\qqbar$ ($q=u,d,s,c$)
production or from semileptonic $B$ decays.  
We reject the $\qqbar$ background by a requirement on a Fisher
discriminant \cite{bib:fisher} ($\Fsfw$) based on a modified set of
Fox-Wolfram moments \cite{bib:fox-wolfram} that characterize the event
topology. 
In the semileptonic $B$ decay background, both $B$ mesons
decay into leptons or two leptons are produced from the $b\to c\to s,d$
decay chain.  
This $B$ decay background is rejected by the requirement on another 
Fisher discriminant variable ($\Fmiss$) constructed from the 
total visible energy $E_{\rm vis} = \sum_i E_i$ and the missing mass 
$M_{\rm miss}
  =\sqrt{(2E_{\rm beam}-\sum E_i)^2-|\sum{\vec{p}_i}|^2}$,  
where $(\vec{p}_i, E_i)$ are the reconstructed 
four-momenta, in the CM frame, of all tracks (assumed to be pions) 
and all photons in the event.

We further reduce the background 
by requiring 
$-0.10\GeV < \Delta E < 0.05\GeV$ for the dielectron channel
($-0.05\GeV < \Delta E < 0.05\GeV$ for the dimuon channel), and 
$\chi^{2}_{\rm vtx}/NDF < 10$.
Here, $\chi^{2}_{\rm vtx}$ is the $\chi^2$ of the $B$ vertex
constructed from the charged daughter particles,
excluding the $K^0_{\rm S}$ daughters.
We also reject candidates with an $\Xs$ invariant mass greater than 
2~$\GeVcc$. This condition removes a large fraction of the combinatorial 
background while retaining 99\% of the signal.

At this stage, there is an average of 1.6 $B$ candidates 
per event in the signal simulation. 
In order to select the most signal-like $B$ candidate, we 
construct a likelihood ratio based on the following six
discriminant variables: 
$\DeltaE$, 
$\DeltaE^{\rm ROE} = E_{\rm ROE}-E_{\rm beam}$, 
$\chi^{2}_{\rm vtx}$,
$\cos\theta_B$,  
$\Fsfw$, and
$\Fmiss$.
The energy of the rest of the event, $E_{\rm ROE}$ is calculated 
by summing the energies of all charged tracks and
neutral calorimeter clusters not included in the $B$
candidate, and 
$\cos\theta_B$ is the cosine of the
$B$ flight direction with respect to the $e^-$ beam
direction in the CM frame.
The variables $\DeltaE$, $\DeltaE^{\rm ROE}$, and $\Fmiss$
are effective at rejecting $\BB$ background, especially
for events with two semileptonic decays, which have large missing energy. 
For continuum suppression, the event-shape variables 
$\Fsfw$ and $\cos\theta_B$ are useful.
The variable $\chi^{2}_{\rm vtx}$ is effective for rejecting 
the random combinatorial background in the high multiplicity modes.

The probability density functions (PDFs) of these
six variables are determined from MC events, separately
for signal and background, with the exception of 
signal PDFs for $\DeltaE$ and $\chi^{2}_{\rm vtx}$. 
We determine the latter two from the charmonium data sample, 
which satisfies all selection criteria except for the charmonium veto
(charmonium-veto sample), 
because we observe some discrepancy between 
the signal MC and the charmonium-veto data sample.
To determine the PDF from data, 
we subtract the background using the shapes obtained
from the $\Mbc$ sidebands. The number of background events
in the $\Mbc$ signal region is normalized by fitting a Gaussian (signal)
and an ARGUS function~\cite{ARGUS} (background) 
to the $\Mbc$ distribution.
For $\DeltaE$, we use separate PDFs 
for the dielectron and dimuon channels.
%


We then calculate the likelihoods 
$\calL_{\rm sig}$ and $\calL_{\rm BG}$ which are products of the 
PDFs of the above six discriminants 
for the signal and the background, respectively.
In each event, only the $B$ candidate with the largest value of the 
likelihood ratio
$\cal R =\calL_{\rm sig}/(\calL_{\rm sig}+\calL_{\rm BG})$
is retained.
According to the MC simulation, this choice implies that 
84\% of the selected $\BtoXsll$ candidates in $\BtoXsll$ 
events have all daughter particles correctly assigned.

\begin{figure}
\begin{center}
\includegraphics[width=16cm]{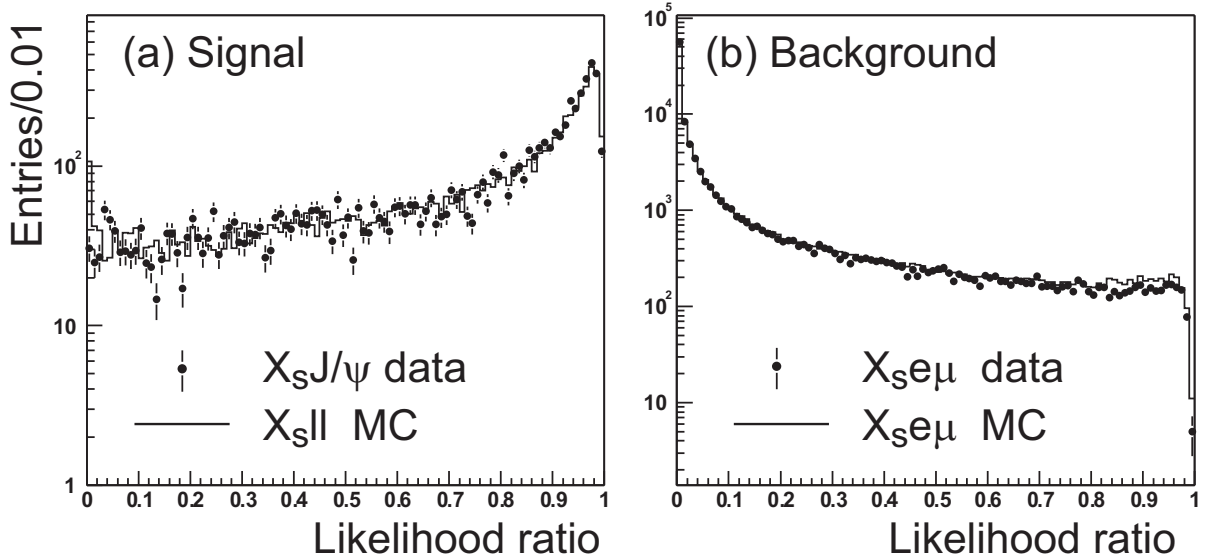}
\caption{
Likelihood ratio for (a) signal and (b) background
events. 
Points represent data and solid histograms
represent MC. Samples used are described in the text.
}
\label{fig:LR_varidation}
\end{center}
\end{figure}

To check the distributions of the likelihood ratio $\cal R$, 
we compare data and MC, separately 
for signal (Fig.~\ref{fig:LR_varidation}a) 
and background (Fig.~\ref{fig:LR_varidation}b) events. 
For the signal, we use charmonium-veto candidates from data and
$\BtoXsll$ signal MC simulation. In the $\cal R$ distribution for
charmonium-veto events, we subtract the background using shapes
obtained from the $\Mbc$ sidebands and 
normalization from the number of background events
in the $\Mbc$ signal region, 
in the same way as for the signal PDF determination.
For the background
we use $\BtoXsemu$ candidates 
that are reconstructed using the nominal selection criteria but
requiring that the two leptons have different flavor,
in data and in a background MC sample that contains 
$b\to c$ decays and $\qqbar$ events. 
We observe good agreement in both cases.

The final suppression of the combinatorial background is
achieved with a requirement on the likelihood ratio ${\cal R}$,
optimized to
maximize the statistical significance of the signal
$S/\sqrt{S+B}$. Here $S$ and $B$ denote the numbers of the signal
and background events, respectively.
This MC optimization is performed separately 
in the regions $M_{X_{s}} < 1.1\GeVcc$ and 
$1.1\GeVcc < M_{X_{s}} < 2.0\GeVcc$, 
resulting in the requirements ${\cal R} > 0.3$ 
and ${\cal R} > 0.9$ respectively.

After applying all selection criteria, 
155 $\BtoXsee$ and 112 $\BtoXsmumu$ candidates 
remain in the $\Mbc$ signal region, defined
as $5.27\GeVcc < M_{\rm bc} < 5.29\GeVcc$.
According to the simulation, the background remaining
at this stage of the analysis consists
mostly of $\BB$ events (80\% and 72\% of the total
background in the electron
and muon channels, respectively). 
In MC signal events, the probability that all daughter 
particles from a selected $\BtoXsll$ candidate are correctly assigned 
is 91\%.

\section{Maximum likelihood fit}
\label{sec:Fit}
We perform an extended unbinned maximum likelihood fit
to the $M_{\rm bc}$ distribution 
in the region $M_{\rm bc} > 5.2\GeVcc$ to extract the signal yield as well as
the shape and yield of the combinatorial background.
The fit is performed separately for the dielectron and dimuon channels.
The likelihood function ${\cal L}$ is expressed as:
\[
{\cal L} = \frac{  e^{-N}} { N!}
           \prod_{ i = 1 }^{N} 
            [(N_{\rm sig} + N_{\rm peak}) {\cal P}^{\rm sig}_{i}
             + N_{\rm pc}   {\cal P}^{\rm pc}_{i} 
	     + N_{\rm cf}   {\cal P}^{\rm cf}_{i} 
	     + N_{\rm comb} {\cal P}^{\rm comb}_{i} 
	    ] 
\]
where $N=N_{\rm sig}+N_{\rm peak}+N_{\rm pc}+N_{\rm cf}+N_{\rm comb}$ 
is the total number of candidate events.
$N_{\rm sig}$, $N_{\rm peak}$, $N_{\rm pc}$, $N_{\rm cf}$, 
and $N_{\rm comb}$ represent the yields of 
the signal, peaking background, combinatorial background from
peaking background, 
cross-feed background (i.e.\ events containing a true $\BtoXsll$
decay, but a selected $B$ candidate with incomplete or wrong decay 
products) 
and combinatorial background,
respectively. 
As explained below, we use the same PDF,
${\cal P}^{\rm sig}_{i}$, 
for the signal and peaking background components.

The signal PDF ${\cal P}^{\rm sig}_{i}$ is described by a Gaussian
for the $\mu^+ \mu^-$ channel as well as for the $e^+ e^-$ channel, since the 
bremsstrahlung recovery and selection procedure for the $e^+ e^-$ channel 
lead to a negligible radiative tail in the $M_{\rm bc}$ distribution.
The Gaussian shape parameters are determined  
from fits of the sum of a Gaussian and an ARGUS function 
to the charmonium-veto data sample.
The fits result in a mean $M_{\rm bc}$ of 
$m_{\rm sig} = 5279.31 \pm 0.05\MeVcc$ and 
a $M_{\rm bc}$ resolution of 
$\sigma_{\rm sig} = 2.62 \pm 0.04\MeVcc$ for the $e^+ e^-$ channel,
and
$m_{\rm sig} = 5279.03 \pm 0.04\MeVcc$ and
$\sigma_{\rm sig} = 2.53 \pm 0.04\MeVcc$ for the $\mu^{+} \mu^{-}$ channel.
In the simulation,
the Gaussian fit results for the $M_{\rm bc}$ distributions of correctly
reconstructed signal events are in agreement with the shape parameters
extracted from the fits to the charmonium-veto sample.
The signal yield $N_{\rm sig}$ is a free parameter in the likelihood fit.

\begin{figure}
\begin{center}
\includegraphics[width=16cm]{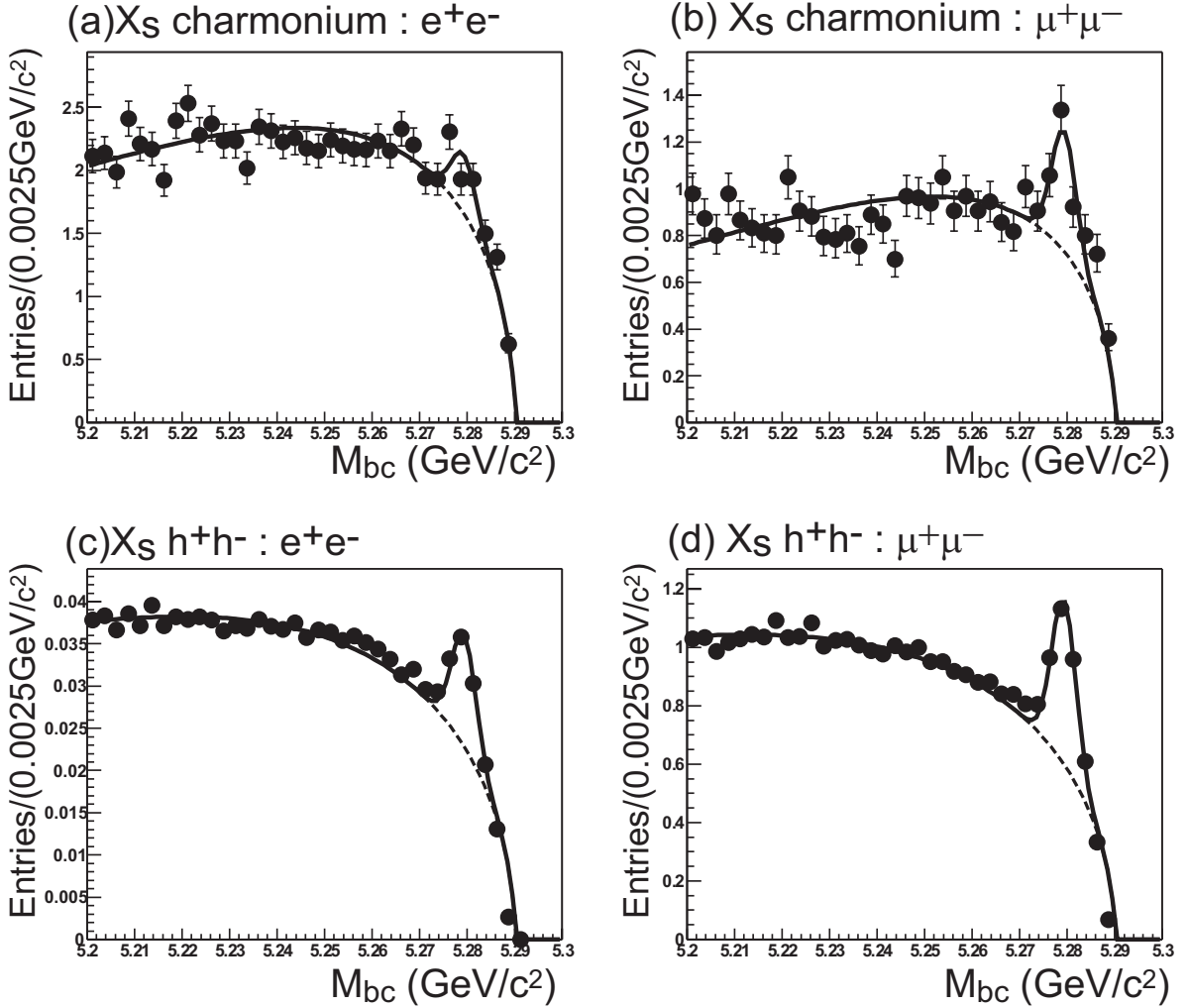}
\caption{
$\Mbc$ distributions 
for MC charmonium events for the (a) dielectron and (b) dimuon channels, 
and for $B \rightarrow X_s h^+h^-$ candidates in data
for the (c) dielectron and (d) dimuon channels.
The normalization corresponds to the expected 
charmonium and hadronic peaking background in $140~\fbinv$ of data.
}
\label{fig:peaking}
\end{center}
\end{figure}
We consider two separate contributions to the peaking background: 
the charmonium peaking background, already described, 
and the hadronic peaking background. The latter arises from the 
$B \to D^{(*)} n \pi$~$(n > 0) \to X_s\pi^+\pi^-$ 
decays where two hadrons 
are misidentified as leptons. 
The normalization and shape of this background 
are determined directly from the data by repeating the 
selection without the lepton identification requirements, 
and fitting the $\Mbc$ distribution with the sum of a Gaussian 
and an ARGUS function. 
We obtain 
$m_{\rm peak} = 5279.16 \pm 0.04\MeVcc$ and
$\sigma_{\rm peak} = 2.60 \pm 0.02\MeVcc$, 
consistent with the signal shape of the charmonium peaking 
background. Hence we use a common PDF for these two peaking 
backgrounds, which is the same as the one determined for the 
signal. Taking into account the momentum and 
angular dependence of the $\pi \to \ell$ misidentification
probabilities (on average 0.08\% for electrons and 0.92\% 
for muons) the hadronic peaking background is estimated to be 
$N_{\rm h-peak} = 0.030 \pm 0.001$ 
events for the dielectron sample and    
$N_{\rm h-peak} = 1.78 \pm 0.05$
events for the dimuon sample.
The charmonium peaking background is estimated from 
the simulation to be
$N_{\rm c-peak} = 1.20 \pm 0.28$ and
$N_{\rm c-peak} = 1.33 \pm 0.21$, respectively. 
In the likelihood fits, the peaking background normalization is 
fixed to $N_{\rm peak} = N_{\rm c-peak} + N_{\rm h-peak}$.
Figures~\ref{fig:peaking}(a) and (b) show the $\Mbc$ distributions 
for the MC charmonium events for the dielectron and dimuon channels, and 
(c) and (d) show the $\Mbc$ distributions for 
$B \rightarrow X_s h^+h^-$ candidates in data 
for the dielectron and dimuon channels, 
respectively. 

The other background PDFs,
${\cal P}^{\rm pc}_{i}$, 
${\cal P}^{\rm cf}_{i}$, and 
${\cal P}^{\rm comb}_{i}$ 
are given by an ARGUS shape.
They describe the
combinatorial contribution from peaking background events,
from cross-feed events, and
from continuum and $\BB$ events, respectively.
The ARGUS cutoff is determined by the 
beam energy in the $\Upsilon(4S)$ rest frame, $E_{\rm beam} = 5.290\GeV$.
The values of the ARGUS shape parameter for each background component
are determined 
from the peaking background $M_{\rm bc}$ distribution shown in 
Fig.~\ref{fig:peaking} (${\cal P}^{\rm pc}_{i}$), 
from incorrectly reconstructed signal MC events
(${\cal P}^{\rm cf}_{i}$), 
and from the fit to the $\BtoXsemu$ data
selected using the nominal selection criteria but
requiring that the two leptons 
have different flavor (${\cal P}^{\rm comb}_{i}$).
We fix these three ARGUS shape parameters.
We also fix the number of peaking background events 
$N_{\rm pc}$ and the number of cross-feed events $N_{\rm cf}$, 
from the peaking background $M_{\rm bc}$ distribution shown in 
Fig.~\ref{fig:peaking} and 
from incorrectly reconstructed signal MC events, 
respectively.
Here the signal MC generations are based on the  
branching fractions in the Ref.~\cite{Ali02}.
The yield $N_{\rm comb}$ is 
taken as a free parameter in the likelihood fit. 

\section{Results}
\label{sec:Results}

\begin{figure}
\begin{center}
\includegraphics[width=16cm]{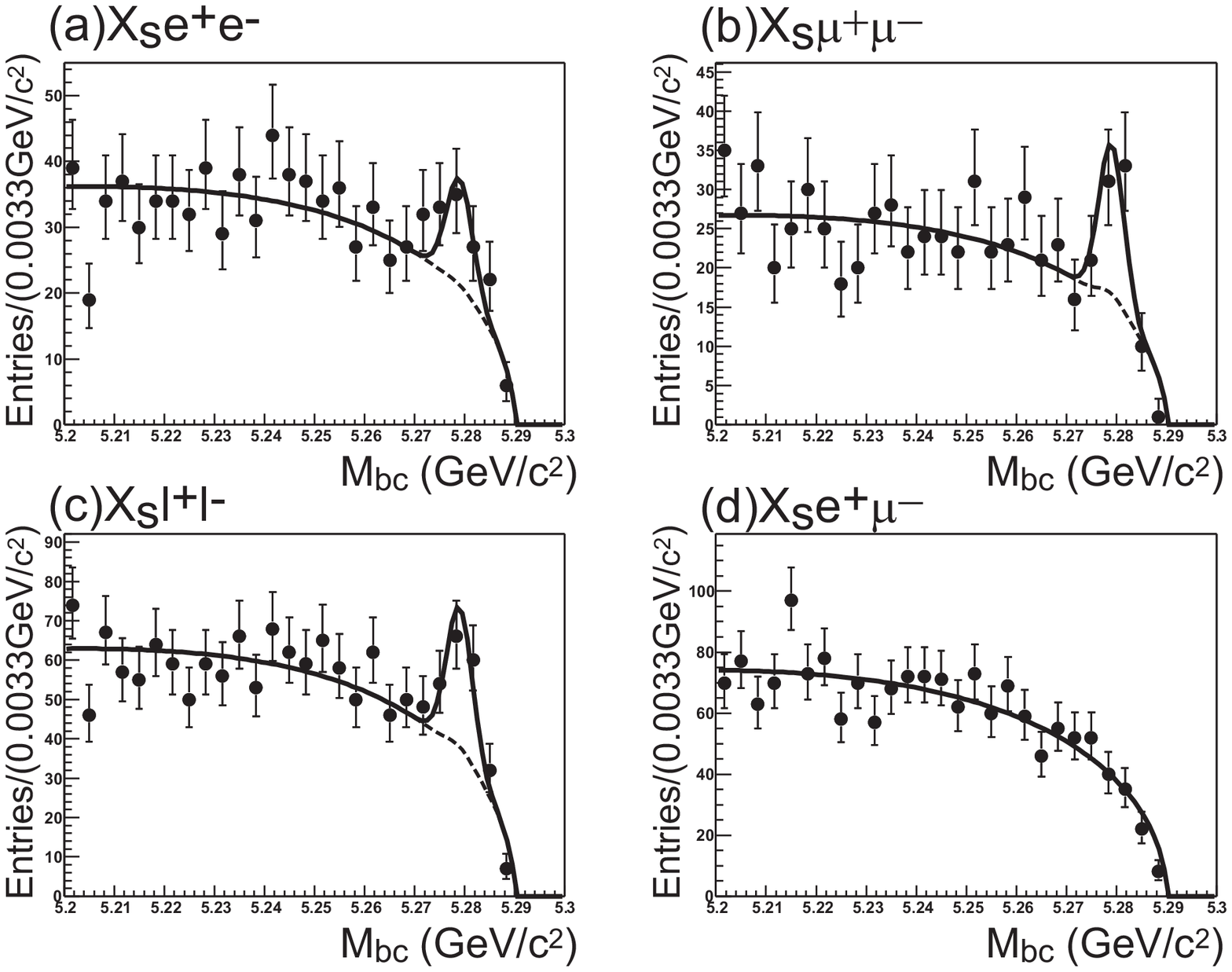}
\caption{$M_{\rm bc}$ distributions of 
(a) $\BtoXsee$, (b) $\BtoXsmumu$, (c) $\BtoXsll$ ($\ell = e, \mu$), 
and (d) $\BtoXsemu$ candidates in data.
The solid lines represent the result of the fits,
and dashed lines represent 
the sum of all background components 
under the signal peaks, respectively.}
\label{fig:fitfig}
\end{center}
\end{figure}

\begin{table}
 \caption{Results of the fit to the data: number of signal candidates
 in the $M_{\rm bc}$ signal box, fitted signal yield $N_{\rm sig}$,
  estimated peaking backgrounds $N_{\rm peak}$ (fixed in the fit),
  and significance including systematics.}
 \begin{center}
 \begin{ruledtabular}
 \begin{tabular}{lcccc}
   Mode & 
       Candidates & $N_{\rm sig}$ & $N_{\rm peak}$ & Significance\\
  \hline
  $X_s\: e^+ e^-$ &    
       155 & $31.8 \pm 10.2$ & $1.24 \pm 0.28$ & \SigBtoXsee \\
  $X_s\: \mu^+\mu^-$ & 
       112 & $36.3 \pm ~9.3$ & $3.11 \pm 0.22$ & \SigBtoXsmumu \\
  Combined &
       267 & $68.4 \pm 13.8$ & $4.35 \pm 0.36$ & \SigBtoXsll \\
 \end{tabular}
 \end{ruledtabular}
\label{tab:fitresult}
 \end{center}
\end{table}

Using the fit parametrization described above, 
we fit the $\Mbc$ distributions in data separately
for the $\BtoXsee$ and $\BtoXsmumu$.
The fit results are displayed in Fig.~\ref{fig:fitfig}
and summarized in Table~\ref{tab:fitresult}.
The significance is
${\cal S} = \sqrt{-2 \ln ({\cal L}^{0}_{\rm max}/{\cal L}_{\rm max})}$,
where ${\cal L}_{\rm max}$ represents the maximum likelihood for the
fit and ${\cal L}^{0}_{\rm max}$ denotes the maximum likelihood 
when the signal yield is constrained to be 0.
To include the systematic uncertainty of the signal yield estimation
into the significance, 
one of the fitting parameters, which are 
signal Gaussian parameters (mean and width), the yield of the peaking BG, 
and the ARGUS shape parameters for the background PDFs 
is varied within $\pm$1$\sigma$ in the above significance calculation.
We then take the minimum significance value.
%
The $\BtoXsll$ signal yield presented in Table~\ref{tab:fitresult} is
obtained by a fit to the combined electron and muon samples.
Figure~\ref{fig:fitfig}(d) shows the $M_{\rm bc}$ distribution for $\BtoXsemu$ 
candidates. 
Applying the ARGUS fit to the $M_{\rm bc}$ distribution, 
there is no evidence for a peaking background as expected.


\begin{figure}
\begin{center}
\includegraphics[width=16cm]{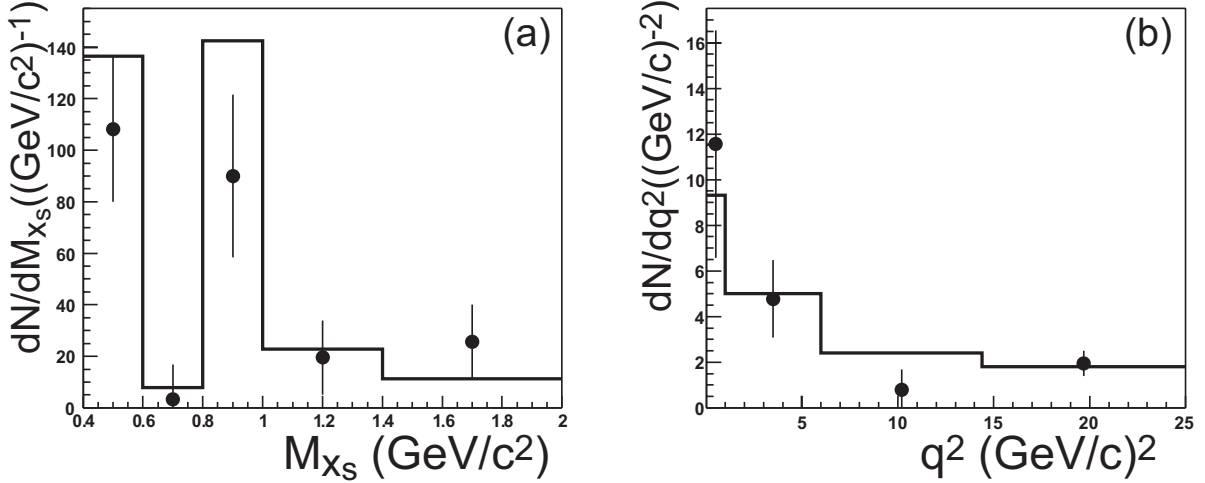}
\caption{Distributions for the signal yield
as a function of (a) hadronic mass $\Mxs$ and (b) $q^2 \equiv \Mll^2$ 
for $\BtoXsll$ signal.
The points with error bars are the data 
(electron and muon channels combined) while the histogram 
is the MC signal normalized to the data. 
}
\label{fig:gra}
\end{center}
\end{figure}

Figures~\ref{fig:gra}(a) and (b) show the distributions of the 
hadronic mass $\Mxs$
and $q^2 \equiv \Mll^2$ for the electron and muon channels combined,
obtained by repeating the likelihood fit in bins of 
$\Mxs$ and $q^2$.
Figure~\ref{fig:gra}(a) indicates that the observed signal includes contributions
from final states across a range of hadronic masses, including
hadronic systems with a mass above that of the $K^\ast(892)$. 


The branching fraction ${\cal B}$ for the signal is calculated as
\begin{equation}
  {\cal B} = \frac{ N_{\rm sig} }{ 2 N_{\BB}~ \epsilon },
\end{equation}
where $N_{\BB} = (152.0 \pm 0.7) \times 10^{6}$ is the number
of $\BB$ pairs produced in $140~\fbinv$ and
$\epsilon$ is the signal efficiency.

\section{Systematic uncertainties}
\label{sec:Systematics}

Systematic uncertainties are of two different
types: those that affect the extraction of the number of signal events
and those that affect the calculation of the branching fraction.
The systematic uncertainties are summarized in Table~\ref{tab_systematics}.

\begin{table}
\caption{
Relative systematic uncertainties (in \%) on the $\BtoXsee$
  and $\BtoXsmumu$ branching fraction measurements. The different 
  contributions are summed in quadrature.}
\begin{center}
\begin{tabular}{lrr}
\hline \hline
Source    	        & $X_s\: e^+e^-$  & $X_s\: \mu^+\mu^-$   \\ \hline\hline
Signal shape                   & $\pm 1.4$       & $\pm 0.5$    \\
BG shape                       & $\pm 7.8$       & $\pm 4.7$    \\
Peaking background statistics  & $\pm 0.9$       & $\pm 0.6$    \\ 
Peaking background PID error   & $<   0.1$       & $\pm 0.5$    \\ 
Peaking background shape       & $\pm 4.3$       & $\pm 2.1$    \\ 
Cross-feed events              & $\pm 4.1$       & $\pm 2.2$    \\ 
\hline
~~~~Signal yield total         & $\pm 9.9$       & $\pm 5.7$    \\ 
\hline
Tracking efficiency            & $\pm 3.5$       & $\pm 3.5$    \\
Lepton identification efficiency    & $\pm 4.1$       & $\pm 5.9$    \\
Kaon identification efficiency      & $\pm 0.8$       & $\pm 0.8$    \\
$\pi^\pm$ identification efficiency & $\pm 0.6$       & $\pm 0.5$    \\
$K^0_{\rm S}$ efficiency    & $\pm 0.7$          & $\pm 0.8$    \\
$\pi^0$ efficiency          & $\pm 0.3$          & $\pm 0.3$    \\
$\cal R$ requirement efficiency& $\pm 5.4$          & $\pm 4.5$    \\ 
Fermi motion model          & $^{+ 6.5}_{-2.4}$  &$^{+ 6.1}_{-2.3}$ \\
${\cal B}(\BtoKll)$         & $\pm 9.9$          & $\pm 10.5$    \\
${\cal B}(\BtoKstarll)$     & $\pm 7.0$          & $\pm 7.8$    \\
$K^\ast$--$X_s$ transition  & $\pm 4.5$          & $\pm 4.7$    \\
Hadronization               & $\pm 8.5$          & $\pm 8.2$    \\
Missing modes               & $\pm 4.5$          & $\pm 4.4$    \\ 
Monte Carlo statistics      & $\pm 1.6$          & $\pm 1.5$    \\
\hline
~~~~Efficiency total 
                            & $^{+19.0}_{-18.1}$ & $^{+19.7}_{-18.9}$   \\
\hline
$\BB$ counting               & $\pm 0.5$          & $\pm 0.5$    \\
\hline\hline
~~~~Total                   & $^{+21.5}_{-20.6}$ & $^{+20.5}_{-19.7}$   \\
\hline\hline
\end{tabular}
\end{center}
\label{tab_systematics}
\end{table}

Uncertainties affecting the extraction of the signal yield
are evaluated by varying the signal Gaussian parameters (mean and width)
and the background shape parameter within $\pm$1$\sigma$ of the
measured values from the charmonium-veto data sample (signal) and 
the $\BtoXsemu$ data sample (background).
We estimate the uncertainties in the peaking-background shape 
and cross-feed events by comparing the signal yields 
obtained with and without the corresponding PDFs 
in the unbinned maximum likelihood fit to $M_{\rm bc}$.

Uncertainties affecting the signal efficiency originate from
the detector modeling, from the simulation of signal decays,
and from the estimate of the number of $B$ mesons in the sample.
By far the largest component is that due to the
simulation of signal decays, discussed in detail below.

The detector modeling uncertainty is composed of the
following uncertainties determined from the data:
the uncertainty in the tracking efficiency of $1.0$\% per track;
the uncertainty in the charged-particle
identification efficiency of $2.05$\% per electron, $2.95$\% per muon,
$1.0$\% per kaon and $0.8$\% per pion; 
and the uncertainty in the reconstruction
efficiency of $4.5$\% per $K^0_{\rm S}$ and $3.3$\% per $\pi^0$.
The efficiency of the likelihood ratio requirement, which suppresses combinatorial
background, is checked with the charmonium-veto sample and the level
of discrepancy with the simulation is taken as the corresponding uncertainty.

The dominant source of uncertainty arises from modeling the signal decays.
Parameters of the Fermi motion model are varied in accordance with
measurements of hadronic moments in semileptonic $B$
decays~\cite{CLEO_moments} and the photon spectrum in inclusive $B \to
X_s\:\gamma$ decays~\cite{CLEO_bsgamma}. 
The fractions of exclusive $\BtoKll$ and $\BtoKstarll$ decays
are varied according to 
theoretical uncertainties~\cite{Ali02}.
The transition point in $M_{X_s}$ between pure $K^\ast \ell^+ \ell^-$
and non-resonant $X_s\:\ell^+\ell^-$ final states is varied
by $\pm 0.1\GeVcc$.

The non-resonant Monte Carlo event generator relies on JETSET to
fragment and hadronize the system consisting of a final
state $s$ quark and a spectator quark from the $B$ meson.
Since the signal efficiencies depend strongly on the
particle content of the final state, uncertainties in the
number of charged and neutral pions and in the 
number of charged and neutral kaons translate
into a significant uncertainty in the signal efficiency
(for $M_{X_s} > 1.1\GeVcc$).

The ratio of the generator yield for decay modes 
containing a $K^0_{\rm S}$ 
to that for modes containing a charged kaon
is varied according to $0.50 \pm 0.11$,
to allow for isospin violation in the decay chain.
The ratio of the generator yield for decay modes 
containing one $\pi^0$ meson 
to that for modes containing none 
is varied according to $1.00 \pm 0.22$.
Uncertainties in these two ratios are set by the
level of discrepancy between $\BtoJpsiX$ data 
and MC $\BtoXsll$ samples.

The 18 modes selected in this analysis only capture about
53\% of the full inclusive rate.
If the contribution of the modes containing a $K^0_{\rm L}$ 
is taken to be equal to that containing a $K^0_{\rm S}$,
the missing states that remain unaccounted for represent $\sim$30\%
of the total rate.
To account for the uncertainty in such remaining missing states, 
we estimate the uncertainty in the fraction of modes
with too many pions or kaons (two extra kaons may be
produced via $s \bar{s}$ popping),
as well as the contributions of modes with photons 
that do not originate from $\pi^0$ decays but rather
from $\eta$, $\eta^\prime$, etc.
For final states with $M_{X_s} > 1.1\GeVcc$, we vary
these fractions by 
$\pm 5\%$ per $\pi^0$, 
$\pm 20\%$ for $\eta$, 
$\pm 30\%$ for $N_{\pi}>5$, and
$\pm 50\%$ for $\eta^{\prime}$ and others.

Including systematic uncertainties, the measured branching fractions
for $M_{\ell^{+} \ell^{-}} > 0.2\GeVcc$ are
\begin{eqnarray}
 {\cal B}(\BtoXsee)   & = & \left(\BrBtoXsee\right)
 \times 10^{-6}, \\
 {\cal B}(\BtoXsmumu) & = & \left(\BrBtoXsmumu\right)
 \times 10^{-6}, \\
 {\cal B}(\BtoXsll)   & = & \left(\BrBtoXsll\right)
 \times 10^{-6},
\end{eqnarray}
where the first error is statistical and the second error is systematic.
The combined $\BtoXsll$ branching fraction is the weighted average of
the branching fractions for the electron and muon channels,
assuming the individual branching fractions to be equal 
for $M_{\ell^{+} \ell^{-}} > 0.2\GeVcc$.
Table~\ref{tab_results} summarizes the results of the analysis
and lists both the statistical and systematic errors on the signal
yields, the signal efficiencies and the branching fractions.

\begin{table}
 \caption{Summary of results: signal yield ($N_{\rm sig}$),
  statistical significance, 
  total signal efficiency $\epsilon$ (including the fraction 
  of $\Xs$ states considered in this analysis)  
  and branching fraction ($\cal B$).
  The first and second errors quoted on $N_{\rm sig}$ and $\cal B$ 
  are statistical and systematic, respectively. The first error on 
  $\epsilon$ corresponds to uncertainties in detector modeling, 
  $\BB$ counting, and Monte Carlo statistics, 
  and the second error on $\epsilon$ to the uncertainties
  in the signal model.
}
 \baselineskip=28pt
 \begin{center}
 \begin{tabular}{lcccc}
  \hline \hline 
  Mode 
    & $N_{\rm sig}$ & Significance & $\epsilon$ (\%) & ${\cal B}~(\times 10^{-6})$ \\
  \hline\hline
  $X_s\: e^+ e^-$
    & $31.8 \pm 10.2 \pm 3.1$ & \SigBtoXsee
    & $\EffBtoXsee $ & $\BrBtoXsee$ \\
  $X_s\: \mu^+\mu^-$
    & $36.3 \pm ~9.3 \pm 2.1$ & \SigBtoXsmumu
    & $\EffBtoXsmumu $ & $\BrBtoXsmumu$ \\
  $X_s\: \ell^+\ell^-$
    & $68.4 \pm 13.8 \pm 5.0$ & \SigBtoXsll
    & $\EffBtoXsll $ & $\BrBtoXsll$ \\
  \hline
 \end{tabular}
 \end{center}
 \label{tab_results}
\end{table}

\begin{figure}
\begin{center}
\includegraphics[width=16cm]{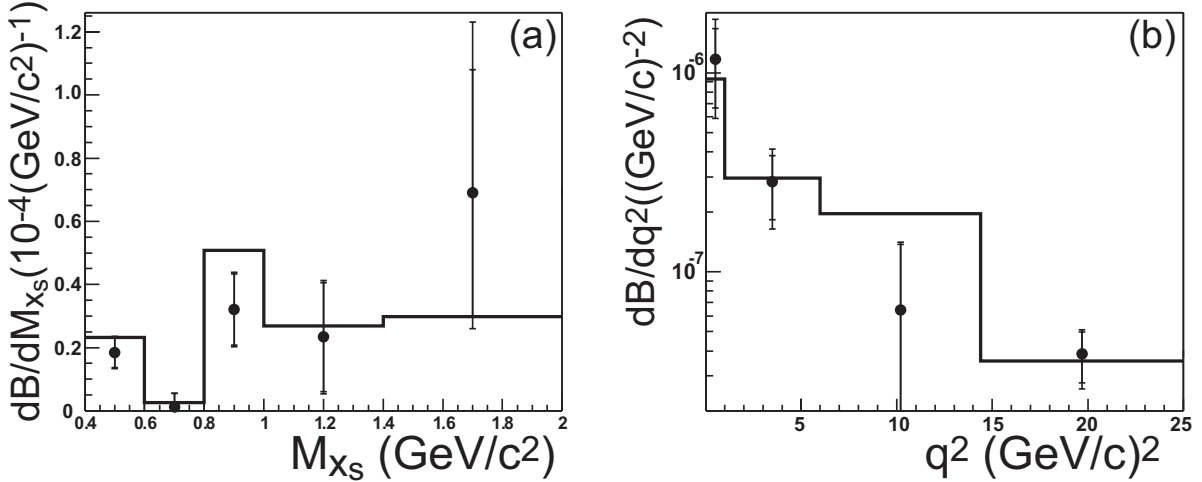}
\caption{Differential branching fraction 
as a function
 of (a) hadronic mass $\Mxs$ and (b) $q^2 \equiv \Mll^2$ 
for $\BtoXsll$ signal.
The points and error bars represent the data 
(electron and muon channels combined) while the histograms 
represent the MC signal normalized to the data statistics. 
The outer (inner) error bars represent the total (statistical)
 errors.}
\label{fig:gra_br}
\end{center}
\end{figure}

The branching fractions for each $\Mxs$ and $q^2$ bin 
are also measured, and summarized in Table~\ref{tab_diffBR}.
Figures~\ref{fig:gra_br}(a) and (b) show the distributions of
the differential branching fractions as a function
 of hadronic mass $\Mxs$ and $q^2 \equiv \Mll^2$, respectively,  
for electron and muon channels combined.

\begin{table}
 \caption{
  $\BtoXsll$ branching fractions ($\cal B$)
  for different bins in $\Mxs$ and $q^2 = M^2_{\ell^+\ell^-}$.
  The first and second errors are statistical 
  and systematic, respectively. 
}
 \baselineskip=28pt
 \begin{center}
 \begin{tabular}{lcclc}
  \hline \hline 
   $\Mxs$ in $\GeVcc$ & ${\cal B}~(\times 10^{-7})$ & \,\,\,\,\, &
   $q^2$ in $({\rm GeV}/c)^{2}$ & ${\cal B}~(\times 10^{-7})$ \\
  \hline
$[0.4,0.6]$  & $3.75 \pm 0.96 ^{+0.34}_{-0.34}$ & &
$[0.04,1.0]$ & $11.34 \pm 4.83 ^{+4.60}_{-2.71}$ \\

$[0.6,0.8]$  & $0.36 \pm 0.88 ^{+0.08}_{-0.08}$ & &
$[1.0,6.0]$  & $14.93 \pm 5.04 ^{+4.11}_{-3.21}$ \\

$[0.8,1.0]$  & $6.65 \pm 2.25 ^{+0.67}_{-0.67}$ & &
$[6.0,14.4]$ & $7.32 \pm 6.14 ^{+1.84}_{-1.91}$ \\

$[1.0,1.4]$  & $10.50 \pm 6.90 ^{+2.00}_{-2.08}$ & &
$[14.4,25.0]$& $4.18 \pm 1.17 ^{+0.61}_{-0.68}$ \\

$[1.4,2.0]$  & $46.59 \pm 23.37 ^{+22.51}_{-10.94}$ & &
     & \\

  \hline

  \hline
  \hline
 \end{tabular}
 \end{center}
 \label{tab_diffBR}
\end{table}

\section{Summary}
\label{sec:Summary}
Using a sample of $152 \times 10^{6}$
$\Upsilon(4S) \to B\overline{B}$ events,
we measure the branching fraction for the rare decay 
$\BtoXsll$, where $\ell = e$ or $\mu$ 
and $\Xs$ is a hadronic system reconstructed semi-inclusively 
in 18 different hadronic states (with up to four pions).
For $M_{\ell^{+} \ell^{-}} > 0.2\GeVcc$, we observe a signal of
$68.4 \pm 13.8(stat) \pm  5.0(syst)$ events
and obtain a branching fraction (averaged over lepton species) of
\[
   \Br(\BtoXsll)=(\BrBtoXsllFull)\times10^{-6},
\]
with a statistical significance of $\SigBtoXsll\:\sigma$.

This result is consistent with the recent prediction
by Ali {\em et al.}~\cite{Ali02,Ali_ICHEP02,Isidori}, 
our previous inclusive $\BtoXsll$ measurement~\cite{Belle_sll}, 
and that of the BaBar collaboration~\cite{BaBar_sll}.


\section*{Acknowledgments}
\label{sec:Acknowledgments}

The authors wish to thank Gudrun~Hiller, Tobias~Hurth and 
Gino~Isidori for their helpful suggestions.
We have also benefited from discussions with Stephane Willocq
regarding Monte Carlo generation with EVTGEN.
We thank the KEKB group for the excellent operation of the
accelerator, the KEK cryogenics group for the efficient
operation of the solenoid, and the KEK computer group and
the National Institute of Informatics for valuable computing
and Super-SINET network support. We acknowledge support from
the Ministry of Education, Culture, Sports, Science, and
Technology of Japan and the Japan Society for the Promotion
of Science; the Australian Research Council and the
Australian Department of Education, Science and Training;
the National Science Foundation of China under contract
No.~10175071; the Department of Science and Technology of
India; the BK21 program of the Ministry of Education of
Korea and the CHEP SRC program of the Korea Science and
Engineering Foundation; the Polish State Committee for
Scientific Research under contract No.~2P03B 01324; the
Ministry of Science and Technology of the Russian
Federation; the Ministry of Education, Science and Sport of
the Republic of Slovenia;  the Swiss National Science Foundation; the National Science Council and
the Ministry of Education of Taiwan; and the U.S.\
Department of Energy.


\end{document}